\documentclass[prb,twocolumn,floatfix,showpacs]{revtex4-1}
\usepackage{graphicx,color}
\usepackage{amsmath,amssymb,bm}

\newcommand{\tn}{\textnormal}
\newcommand{\cprb}[3]{Phys.~Rev.~B {\bf #1}, #2 (#3)}
\newcommand{\cprl}[3]{Phys.~Rev.~Lett.~{\bf #1}, #2 (#3)}

\newcommand{\cbook}[2]{\textit{#1} (#2)}

\definecolor{darkred}{rgb}{0.90,0,0}
\definecolor{darkgreen}{rgb}{0,0.60,.2}
\definecolor{darkblue}{rgb}{0,0,1}
\definecolor{grey}{cmyk}{0,0,0,0.25}
\definecolor{orange}{cmyk}{0,0.6,0.8,0}

\begin{document}

\title{Transport properties of the one-dimensional Hubbard model at finite temperature}

\author{C.\ Karrasch$^{1,2}$}
\author{D.\ M.\ Kennes$^{3}$}
\author{J.\ E.\ Moore$^{1,2}$}

\affiliation{$^1$Department of Physics, University of California, Berkeley, California 95720, USA}
\affiliation{$^2$Materials Sciences Division, Lawrence Berkeley National Laboratory, Berkeley, CA 94720, USA}
\affiliation{$^3$Institut f\"ur Theorie der Statistischen Physik, RWTH Aachen University and JARA---Fundamentals of Future Information Technology, 52056 Aachen, Germany}

\begin{abstract}

We study finite-temperature transport properties of the one-dimensional Hubbard model using the density matrix renormalization group. Our aim is two-fold: First, we compute both the charge and the spin current correlation function of the integrable model at half filling. The former decays rapidly, implying that the corresponding Drude weight is either zero or very small. Second, we calculate the optical charge conductivity $\sigma_\tn{reg}(\omega)$ in presence of small integrability-breaking next-nearest neighbor interactions (the extended Hubbard model). The DC conductivity is finite and diverges as the temperature is decreased below the gap. Our results thus suggest that the half-filled, gapped Hubbard model is a normal charge conductor at finite temperatures. As a testbed for our numerics, we compute $\sigma_\tn{reg}(\omega)$ for the integrable XXZ spin chain in its gapped phase.

\end{abstract}

\maketitle



\section{Introduction}

The physics of one-dimensional (1D) systems is strongly influenced by electronic correlations. E.g., the low-energy behavior of a large class of 1D models is not described by a Fermi liquid but exhibits bosonic excitations. Thermodynamic properties or equilibrium correlation functions of this so-called Luttinger liquid can be obtained elegantly using field theory. Transport properties, however, are usually not governed by the low-energy Luttinger liquid fixed point but by an interplay between dangerously irrelevant operators scattering the currents and conserved quantities protecting them.\cite{andrei,xxz_sirker,xxz_prosen} In order to connect to actual experimental transport measurements on (quasi) 1D systems such as carbon nanotubes or strongly anisotropic 3D materials, it is thus essential to study generic microscopic models.

The Hubbard model\cite{hubbook} plays a fundamental role in the physics of correlated electrons as it provides the most transparent realization of Mott physics: it describes spinful electrons moving on a lattice with on-site interaction. From an experimental perspective, the Hubbard model has been employed to describe a wide variety of crystals (both insulating and conducting); in one dimension, it is also used as a starting point for polymers. Despite the Hubbard model's apparent simplicity and the existence of an exact solution for its thermodynamics in 1D, little is known about its transport properties. Above one dimension even the phase diagram of the Hubbard model with doping is a matter of debate, including the regime of greatest interest where charged excitations are gapped while the neutral sector includes gapless spin excitations. 

From a theoretical perspective, it is less demanding (both computationally and analytically) to study a \textit{spinless} fermion model with nearest-neighbor interactions, which has therefore been investigated extensively despite the fact that it is of less relevance for experimental setups. Even though the model (which can be mapped to a XXZ spin chain via a Jordan-Wigner transformation) allows for a Bethe ansatz solution which is `simpler' than that of the Hubbard model, extracting transport coefficients, which are determined by couplings between all excitations, remains a formidable task. Over the last decades a significant number of works\cite{bethe0,xxz_betheT1,xxz_betheT2,xxz_qmcgros,xxz_edmillis,xxz_qmcsorella,xxz_fieldtheory,xxz_fabian,xxz_rosch,xxz_edherbrych,xxz_sirker} investigated the equilibrium transport properties of the XXZ chain, but it was only recently proven rigorously\cite{xxz_prosen,xxz_prosen2} that the half-filled system can support dissipationless currents at finite temperature $T>0$; this corresponds to a finite Drude contribution to the conductivity
\begin{equation}\label{eq:sigma}
\sigma(\omega) = 2\pi D \delta(\omega) + \sigma_\tn{reg}(\omega)\,.
\end{equation}
Numerical values for the Drude weight $D$ can be obtained, e.g., via the density matrix renormalization group.\cite{drudepaper,drudepaper2}

It is the main goal of this paper to investigate finite-temperature linear-response transport properties of the Hubbard model, about which comparably little is known (we will give a more detailed overview of previous works in Sec.~\ref{sec:dw}). In particular, we employ the density matrix renormalization group\cite{white1,dmrgrev,dmrgrev2} to compute real time charge- and spin current correlation functions whose Fourier transform determines $\sigma(\omega)$. Our focus is two-fold: First, we demonstrate that the charge current correlators at half filling decay rapidly, suggesting that the charge Drude weight $D^c$ is either zero or very small at any finite temperature ($D^c$ is known to vanish at $T=0$). Second, we study the optical conductivity $\sigma_\tn{reg}(\omega)$ in presence of small next-nearest neighbor interactions which break integrability\cite{integrab1,integrab2} -- and thus eliminate a potentially small, finite Drude weight -- but do not trigger a phase transition.\cite{extendedhub} As the temperature is lowered below the charge gap, $\sigma_\tn{reg}(\omega)$ successively develops a sharp increase at the optical absorption threshold. More importantly, the DC conductivity $\sigma_\tn{reg}(0)$ is finite and diverges approximately as $1/T$ (our data is insufficient to rule out a weakly $T$-dependent prefactor). This analysis suggests that the half-filled, gapped Hubbard model is a normal charge conductor at finite temperatures. As a testbed for our numerics, we also study the integrable XXZ chain in its gapped phase.\cite{diff_zotos,diff_prosenznidaric,diff_steinigeweggemmer,diff_steinigeweg1,diff_znidaric,diff_znidaric2,diff_steinigeweg2,finiteTquench,diff_typicality} The DC conductivity again grows monotonously for all temperatures from $T=\infty$ to far below the gap.

\section{Model and Method}
\label{sec:model}

\textit{Models} --- The prime interest of this work is the extended Hubbard model governed by
\begin{equation}\begin{split}
H =& \sum_{l=1}^{L-1}\Big\{-\sum_\sigma \Big[\frac{1}{2}  c_{l\sigma}^\dagger c_{l+1\sigma}^{\phantom{\dagger}} + \tn{h.c.} \Big]
+ Un_{l\uparrow}n_{l\downarrow} \\
&+V\big( n_{l\uparrow}+n_{l\downarrow}\big)\big( n_{l+1\uparrow}+n_{l+1\downarrow}\big)\Big\}\,,
\end{split}\end{equation}
with $n_{l\sigma}= c_{l\sigma}^\dagger c_{l\sigma}^{\phantom{\dagger}}-1/2$, and $c_{l\sigma}$ being fermionic annihilation operators. We solely focus on the case of half filling and zero magnetic field (the reason for this will be outlined below). At $V=0$, the model is symmetric under $U\to-U$ if charge and spin degrees of freedom are interchanged, and we thus stick to repulsive interactions $U\geq0$ only. The Hubbard model is integrable via Bethe ansatz for $V=0$. A charge gap opens for $U>0$ while the spin sector remains gapless.\cite{hubbook}

The second model we study is the XXZ chain defined by
\begin{equation}
H = \sum_{l=1}^{L-1}\left[ S^x_lS^x_{l+1} + S^y_lS^y_{l+1} + \Delta S^z_lS^z_{l+1}\right]\,,
\end{equation}
where $S^{x,y,z}_l$ are spin-$1/2$ operators. The model is integrable via Bethe ansatz; a gap opens for $|\Delta|>1$. At $U\to\infty$ and $V=0$, the Hubbard model can be mapped to an isotropic, antiferromagnetic XXZ chain ($\Delta=1$).

\textit{Transport coefficients} --- Both the Drude weight $D$ and the regular part $\sigma_\tn{reg}$ of the conductivity of Eq.~(\ref{eq:sigma}) can be obtained from the current correlation function,
\begin{equation}\label{eq:dw}
D = \lim_{t\to\infty}\lim_{L\to\infty} \frac {\textnormal{Re } \langle J(t)J(0)\rangle}{2LT}\,,
\end{equation}
and
\begin{equation}\label{eq:sigma1}\begin{split}
\tn{Re}\,\sigma_\tn{reg}(\omega) =\, & \frac{1-e^{-\omega/T}}{\omega}\times \\
& \tn{Re} \int_0^{\infty}dte^{i\omega t} \lim_{L\to\infty} \frac{\langle J(t)J(0)\rangle_\tn{reg}}{L}\,,
\end{split}\end{equation}
where a potentially finite Drude weight has been subtracted in $\langle J(t)J(0)\rangle_\tn{reg}$. Only finite times can be reached in the DMRG and it is thus imperative to estimate the associated error of $\sigma_\tn{reg}(\omega)$. In the DC limit $\omega=0$ where `finite-time' effects are largest, only $\tn{Re}\,\langle J(t)J(0)\rangle$ contributes to Eq.~(\ref{eq:sigma1}), and one can therefore estimate the error by comparing with $\sigma_\tn{reg}$ obtained from $\tn{Im}\,\langle J(t)J(0)\rangle$,
\begin{equation}\label{eq:sigma2}
\tn{Re}\,\sigma_\tn{reg}(\omega) = -\frac{2}{\omega}\,
 \tn{Im} \int_0^{\infty}dte^{i\omega t} \, \tn{Im}\lim_{L\to\infty} \frac{\langle J(t)J(0)\rangle_\tn{reg}}{L}\,,
\end{equation}
which follows from a Kramers-Kronig relation.

The current operator $J = \sum_l j_l$ is defined via a continuity equation. The local charge and spin current of the Hubbard model read
\begin{equation}
j_{l}^\tn{c,s} = i\big( c_{l\uparrow}^\dagger c_{l+1\uparrow}^{\phantom{\dagger}} - c_{l+1\uparrow}^\dagger c_{l\uparrow}^{\phantom{\dagger}}\big)
\pm i\big( c_{l\downarrow}^\dagger c_{l+1\downarrow}^{\phantom{\dagger}} - c_{l+1\downarrow}^\dagger c_{l\downarrow}^{\phantom{\dagger}}\big)\,,
\end{equation}
and for the XXZ chain one finds
\begin{equation}
 j_{l} = i\big(S^x_lS^y_{l+1} - S^y_lS^x_{l+1} \big)\,.
\end{equation}

\begin{figure}[t]
\includegraphics[width=0.95\linewidth,clip]{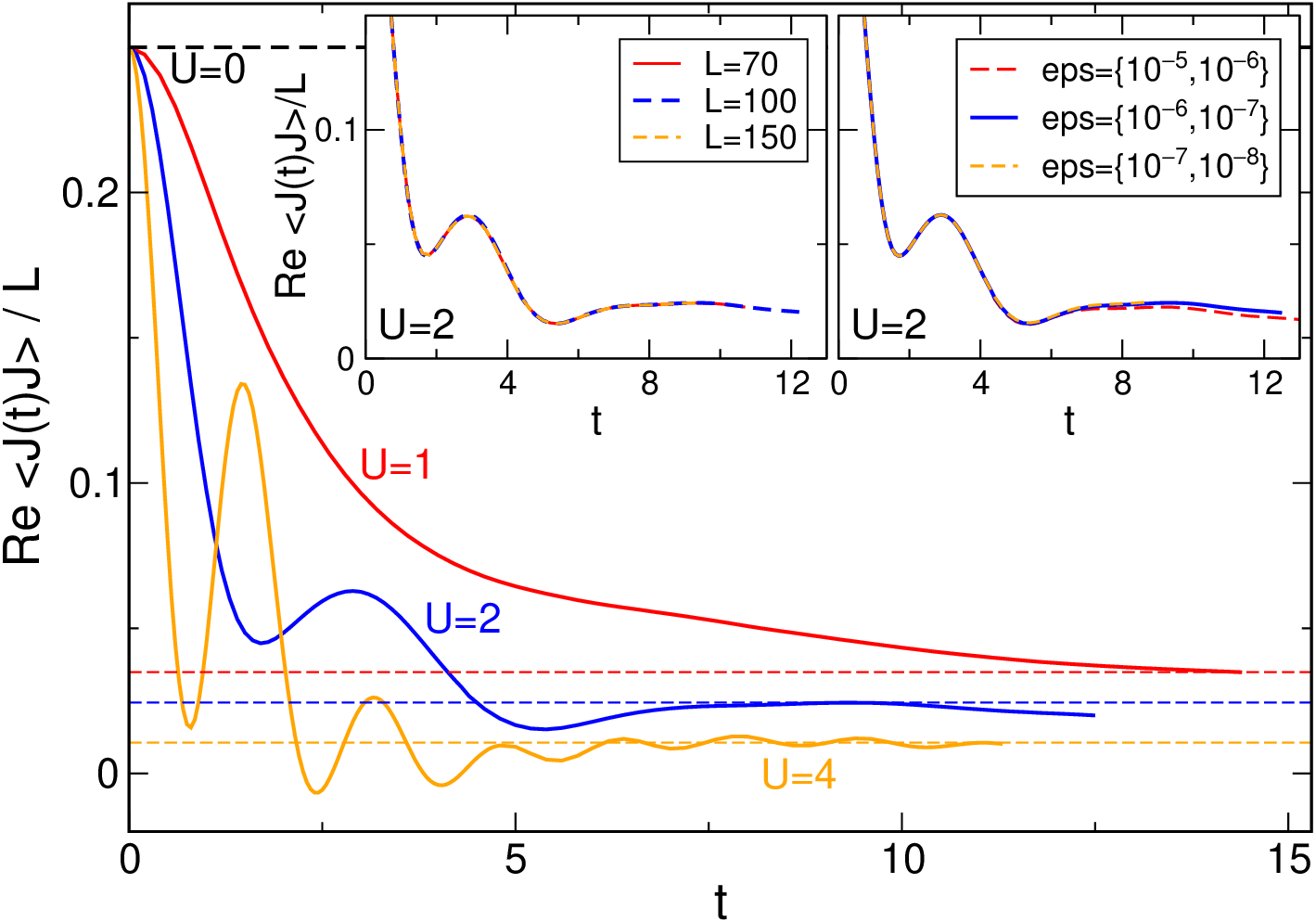}
\caption{(Color online) Current correlation function of the integrable Hubbard model at half filling, $L=100$, and infinite temperature $T=\infty$ (where charge and spin degrees of freedom are symmetric). The long-time asymptote determines the Drude $D$ weight via Eq.~(\ref{eq:dw}). The numerics indicate that $\lim_{T\to\infty}[TD]$ is zero or small; an estimate for upper bounds is given by the dashed lines. \textit{Insets:} DMRG data for various system sizes $L$ and discarded weights $\{\epsilon_j,\epsilon_J\}$ during the real time evolution of the local $j$ and global $J$, respectively. }
\label{fig:dw_inf}
\end{figure}

\begin{figure*}[t]
\includegraphics[width=0.475\linewidth,clip]{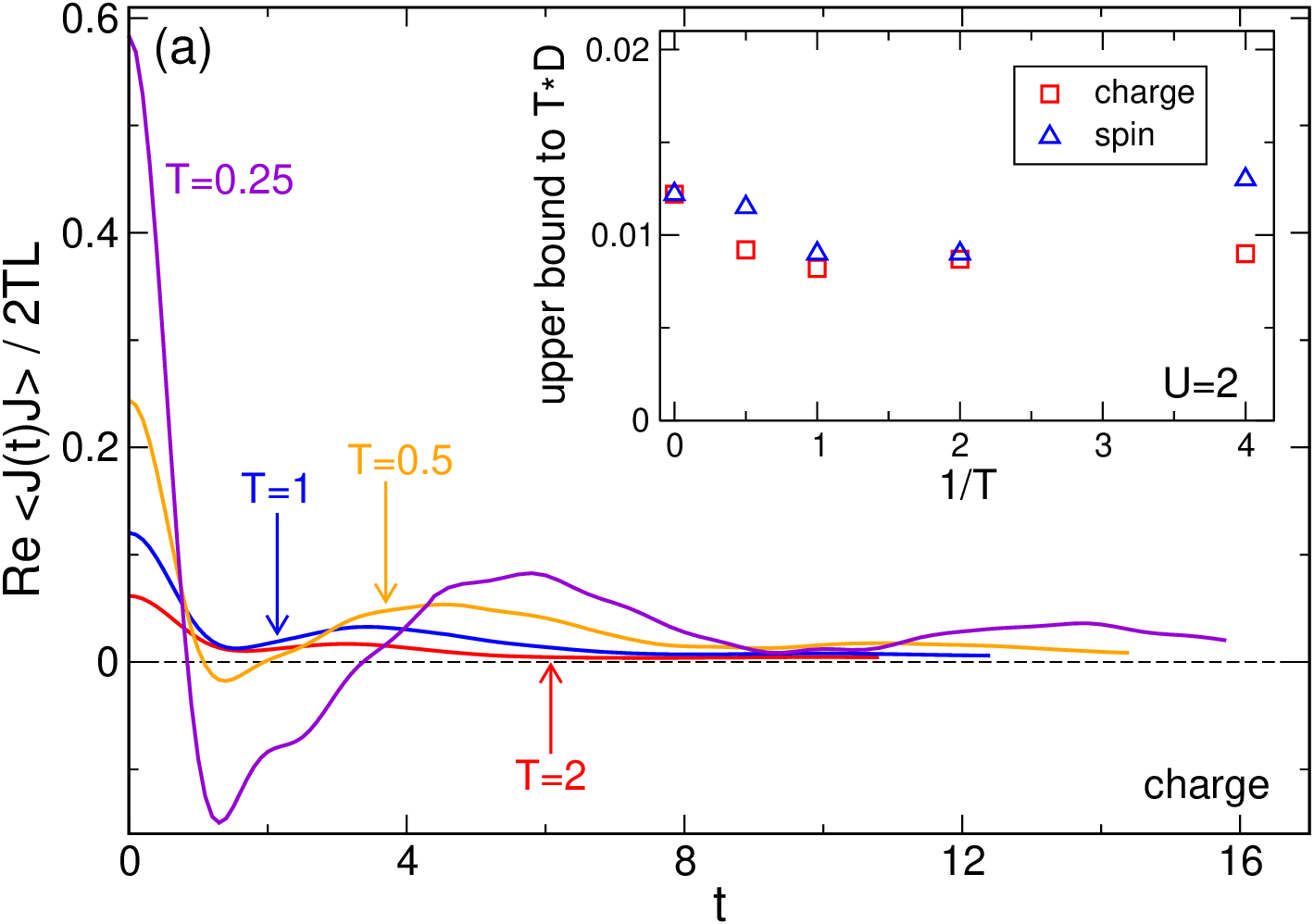}\hspace*{0.03\linewidth}
\includegraphics[width=0.475\linewidth,clip]{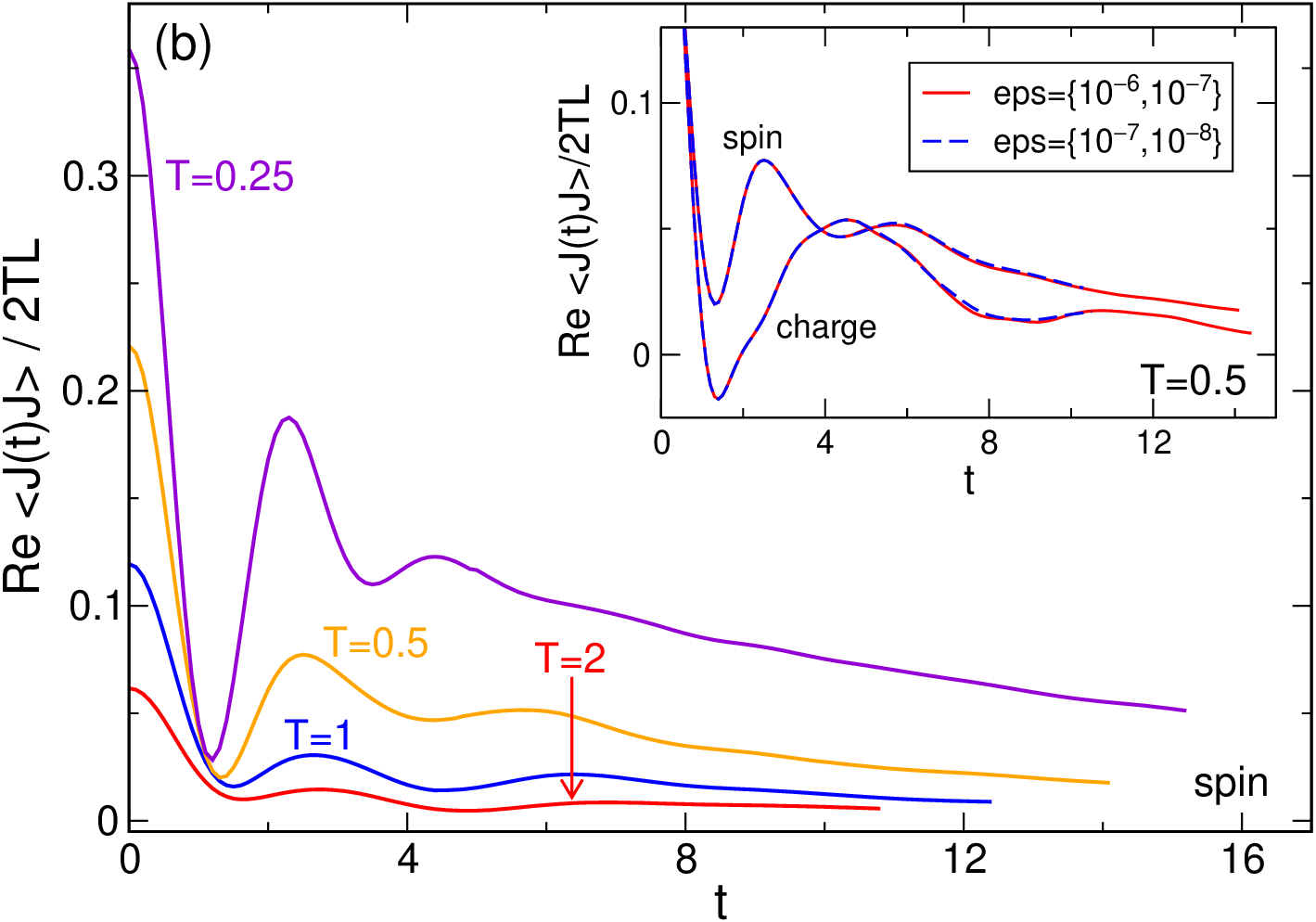}
\caption{(Color online) The same as in Fig.~\ref{fig:dw_inf} but for various finite temperatures. Panels (a) and (b) show the charge and spin current correlation function, respectively. \textit{Inset to (a)}: Upper bound for the Drude weights $TD=\lim_{t\to\infty}\lim_{L\to\infty}\tn{Re}\langle J(t)J(0)\rangle/L$ at $U=2$. \textit{Inset to (b)}: DMRG data for various discarded weights.}
\label{fig:dw_U2}
\end{figure*}

\textit{DMRG} --- We compute the current correlation function
\begin{equation}\label{eq:time}
 \langle J(t) J\rangle \sim \tn{Tr}\big[ e^{-H/T}e^{iHt}Je^{-iHt}J\big]
\end{equation}
using the time-dependent\cite{tdmrg1,tdmrg2,tdmrg3,tdmrg4,tdmrg5,tdmrg6} density matrix renormalization group\cite{white1,dmrgrev,dmrgrev2} in a matrix product state\cite{mps1,mps2,mps3,mps4} implementation. Finite temperatures\cite{dmrgT,barthel,verstraete,vidalop,tmrg1,metts,trick2a} are incorporated via purification of the thermal density matrix. The real- and imaginary time evolution operators in Eq.~(\ref{eq:time}) are factorized by a fourth order Trotter decomposition with a step size of $dt=0.05\ldots0.2$. The discarded weight during each individual `bond update' is kept beneath a threshold value $\epsilon$, which is the key parameter controlling the accuracy of the simulation. The system size, however, can easily be chosen large enough for the results to be effectively in the thermodynamic limit. We carry out most of our calculations for $L=100$ or $L=200$ and exemplary compare against other values. The dependence of the numerical data on $L$ and $\epsilon$ is illustrated in the insets to Figs.~\ref{fig:dw_inf} and \ref{fig:dw_U2}(b).

The bond dimension $\chi$ increases exponentially fast during the real time evolutions, and the simulation is stopped once numerical resources are exhausted. We pursue several strategies in order to access time scales as large as possible. First, we employ the finite-temperature disentangler introduced in Ref.~\onlinecite{drudepaper}, which uses the fact that purification is not unique to slow down the growth of $\chi$. Second, we `exploit time translation invariance',\cite{trick2a} rewrite $\langle J(t)J(0)\rangle=\langle J(t/2)J(-t/2)\rangle$, and carry out two independent calculations for $J(t/2)$ as well as $J(-t/2)$. This allows to access time scales roughly twice as large. Third, for the XXZ chain we recast $\langle J(t)J(-t)\rangle=2L \langle J^\uparrow(t)j_{L/2}(-t)\rangle$ with $J^\uparrow=i\sum_lS^x_lS^y_{l+1}$, and similarly for the charge and spin currents of the Hubbard model. We exploit $U(1)$ symmetries (e.g., both charge and spin conservation) during the time evolution and stop the calculation once the bond dimension reaches $\chi_j\sim3000$ for the local $j_{L/2}$ and $\chi_J\sim 1500$ for the global $J$ ($\chi_j\sim4000$, $\chi_J\sim3000$ in some exemplary cases).

\section{Drude weight}
\label{sec:dw}

We start this section with a brief summary of what is know about the Drude weight of the Hubbard model. For $U=V=0$ (free fermions), both the spin and charge currents are conserved by $H$, and thus trivially $D^{c,s}(T\geq0)>0$. At $U>0$ but $V=0$, the model is integrable via Bethe ansatz,\cite{hub0} and away from half filling it follows from the Mazur inequality that the Drude weights are finite at any temperature.\cite{integrab2} At half filling, however, all known local conserved quantities\cite{hub1} have zero overlap with the current operators for symmetry reasons. At $T=0$ one can use Bethe ansatz\cite{bethe0} to show that the charge Drude weight $D^c$ vanishes while the spin Drude weight $D^s$ is finite (recall that the charge sector is gapped while the spin sector is gapless). Further studies of ground state transport properties can be found in Refs.~\onlinecite{hubT0a,hubT0b,hubT0c,hubT0d,hubT0e,hubT0f,hubT0g,hubT0h,hubT0i,hubT0j,hubT0k,hubT0l,hubT0m,hubT0n}. 

\begin{figure}[b]
\includegraphics[width=0.95\linewidth,clip]{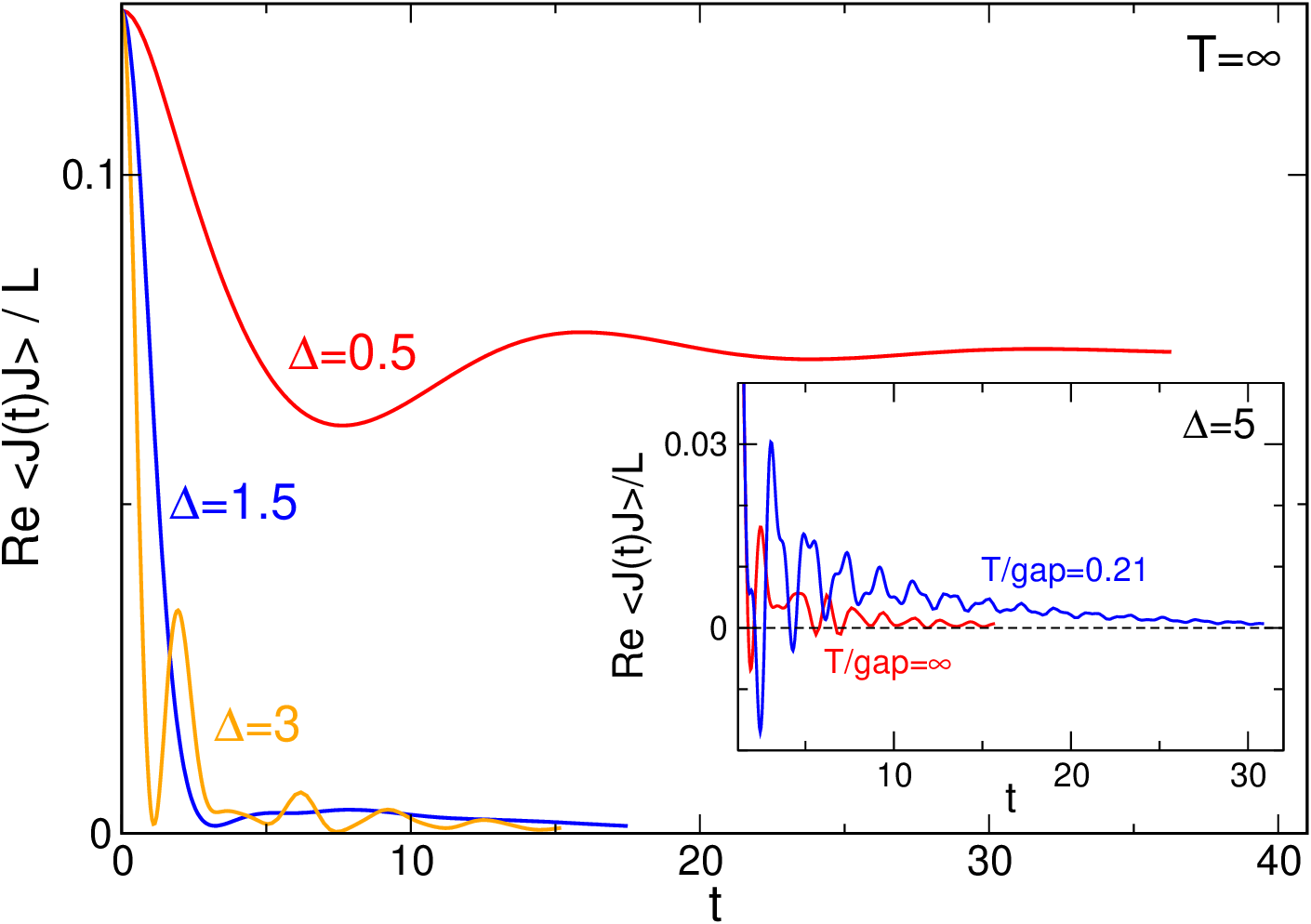}
\caption{(Color online) Current correlation function of the XXZ spin chain, illustrating that the Drude weight is finite in the gapless phase ($|\Delta|<1$) but vanishes in the gapped phase. The spectral gap $\delta$ is know exactly from Bethe ansatz: $\delta=0.087$ at $\Delta=1.5$, $\delta=1.23$ at $\Delta=3$, and $\delta=3.12$ at $\Delta=5$.}
\label{fig:dw_xxz}
\end{figure}

\begin{figure*}[t]
\includegraphics[width=0.475\linewidth,clip]{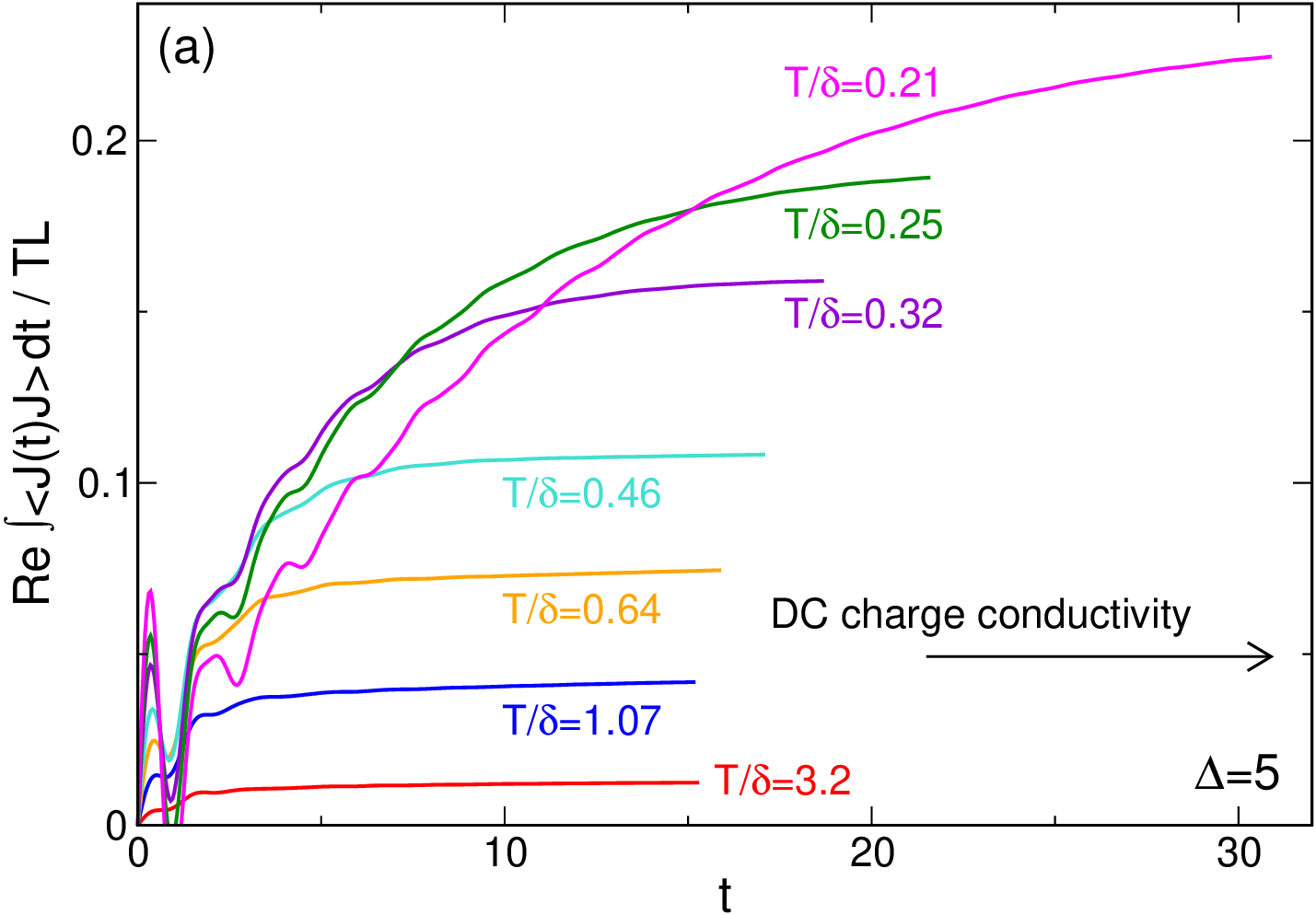}\hspace*{0.03\linewidth}
\includegraphics[width=0.475\linewidth,clip]{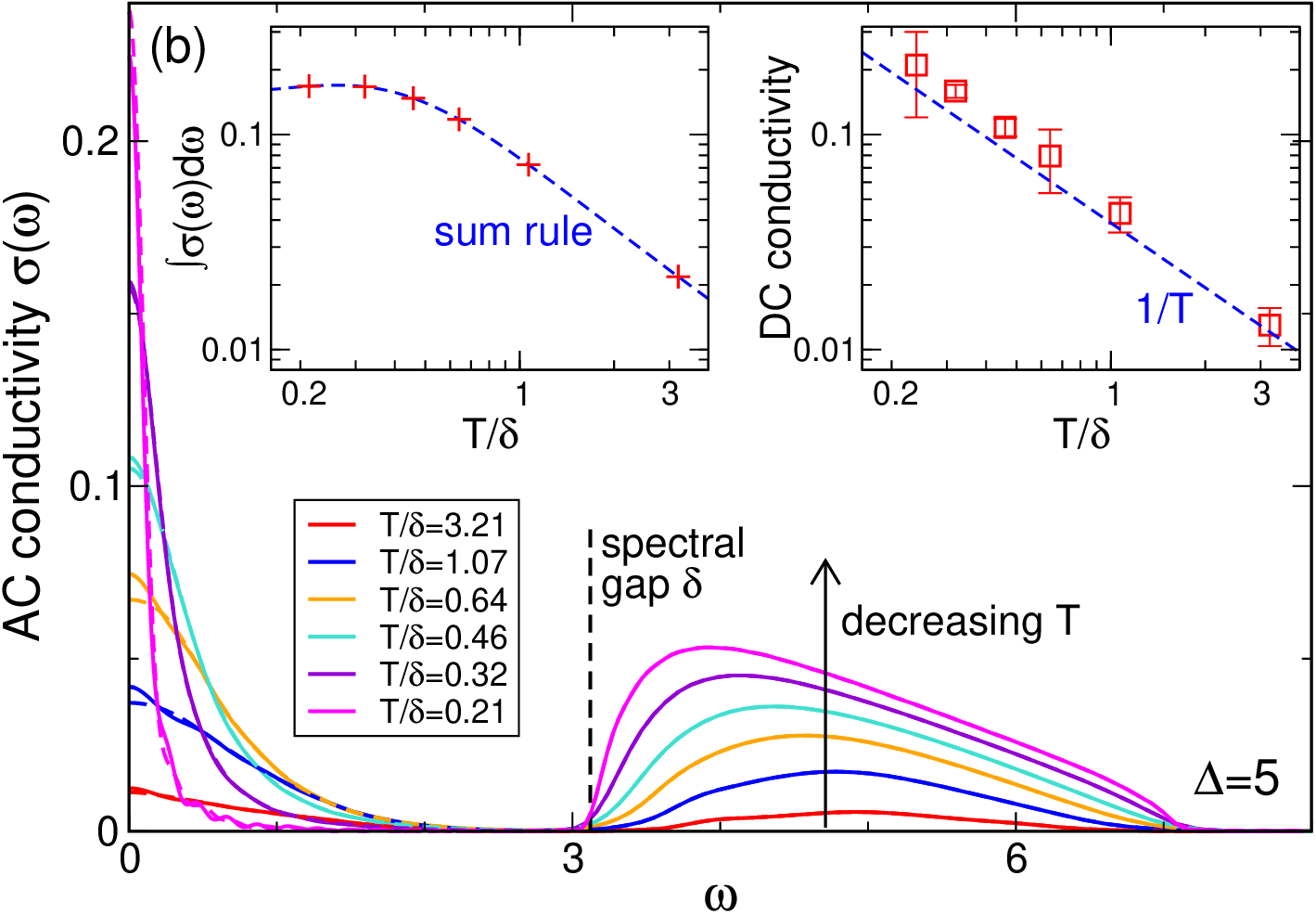}
\caption{(Color online) (a) Integral of the current correlation function of the XXZ chain in the gapped phase with $\Delta=5$. The Drude weight vanishes, and the DC conductivity is given by the long-time asymptote of the curves. It increases as the temperature is lowered from $T=\infty$ to below the spectral gap $\delta=3.12$. (b) Optical conductivity obtained via Eq.~(\ref{eq:sigma1}) (solid lines) or via Eq.~(\ref{eq:sigma2}) (dashed lines, almost indistinguishable) without extrapolation of the finite-time data. \textit{Left Inset:} The sum rule $\int_0^\infty\sigma(\omega)d\omega=-\pi\langle H_\tn{kin}\rangle/2L$ is fulfilled with great accuracy ($\langle H_\tn{kin}\rangle$ can be obtained easily from infinite-system DMRG). \textit{Right Inset:} Conductivity in the DC limit (see the main text for an explanation of the error bars).}
\label{fig:sigma_xxz}
\end{figure*}

The most interesting situation is thus the half-filled Hubbard model at finite temperature, where there is still controversy about whether or not the Drude weight is finite. In Refs.~\onlinecite{integrab1,integrab2} it was conjectured that for an integrable model $D(T)>0$ if and only if $D(T=0)>0$, which holds true for the XXZ chain whose finite-$T$ Drude weight is non-zero in the gapless phase but vanishes in the gapped phase. For the Hubbard model, Bethe ansatz results\cite{betheThub} suggest that $D^c(T>0)>0$ despite the charge gap. However, this calculation is very involved: two different ways\cite{xxz_betheT1,xxz_betheT2} to approximately solve the Bethe ansatz equations for the XXZ chain yield results which disagree, and one might expect the situation for the Hubbard model to be similarly subtle. While quantum Monte Carlo (QMC) numerics\cite{qmchub} and an analysis of low-energy excitations\cite{exhub} support a finite Drude weight, exact diagonalization (ED) data,\cite{edhub} large-$U$ analytics,\cite{largeUhub} and symmetry arguments\cite{symhub} seems to favor $D^c=0$. Non-equilibrium DMRG calculations (at $T=\infty$ only) suggest a zero Drude weight but a finite diffusion constant $\lim_{T\to\infty}T\sigma_\tn{reg}(0)$.\cite{prosenhub} The situation for the Hubbard model is thus completely analogous to the gapless XXZ chain where -- after decades of dispute -- it was only recently shown rigorously that the Drude weight at half filling and $\Delta\neq1$ is nonzero at finite $T$.\cite{xxz_prosen,xxz_prosen2,xxz_prosenD1}

The controversial status of the Drude weight of the half-filled Hubbard model requires further work in this direction. The DMRG allows to obtain results directly on the real time axis (in contrast to QMC), and the thermodynamic limit can be accessed easily (in contrast to ED). Its drawback is that only finite time scales can be reached. We show DMRG results for the current correlation functions of the half-filled, integrable Hubbard model at $T=\infty$ (where spin and charge are symmetric) in Fig.~\ref{fig:dw_inf}. Finite temperatures are shown in Fig.~\ref{fig:dw_U2}. The charge current correlation functions fall off rapidly at any $T$, suggesting that the charge Drude weight vanishes at finite temperatures in agreement with Refs.~\onlinecite{edhub,largeUhub,symhub} but in contrast to the Bethe ansatz\cite{betheThub} and QMC\cite{qmchub} predictions. However, the times reached in our numerics are too small to unambiguously rule out a small but finite $D^c>0$. It is thus instructive to establish an \textit{upper} bound to $D^c$ as a reference for future works. Such a bound can be defined from the value of $\langle J(t)J\rangle$ at the largest time in case that it falls off monotonously or from the value at the last maximum in case $\langle J(t)J\rangle$ oscillates (this is illustrated by the dashed lines in Fig.~\ref{fig:dw_inf}). The upper bound for the charge Drude weight at finite $T$ and $U=2$ is shown in the inset to Fig.~\ref{fig:dw_U2}(a). It is significantly \textit{smaller} than the values estimated by QMC.\cite{qmchub}

In contrast, the spin current correlation functions decay on an increasingly larger time scales as the temperature is decreased. This is consistent with the spin Drude weight being nonzero at $T=0$.\cite{bethe0} At infinite temperature, however, spin and charge degrees of freedom are symmetric, $\lim_{T\to\infty}TD^c=\lim_{T\to\infty}TD^s$. If we \textit{assume} that $TD^c$ is zero at all $T$, this leaves two possible scenarios for the temperature dependence of $D^s$: either $D^s$ is nonzero at any finite $T$ but vanishes in the limit of $T\to\infty$ faster than $1/T$, or $D^s(T>0)=0$. Our numerical data is insufficient to answer this question conclusively. For reasons of completeness, an upper bound to $D^s$ is shown in the inset to Fig.~\ref{fig:dw_U2}(a).

\begin{figure*}[t]
\includegraphics[width=0.475\linewidth,clip]{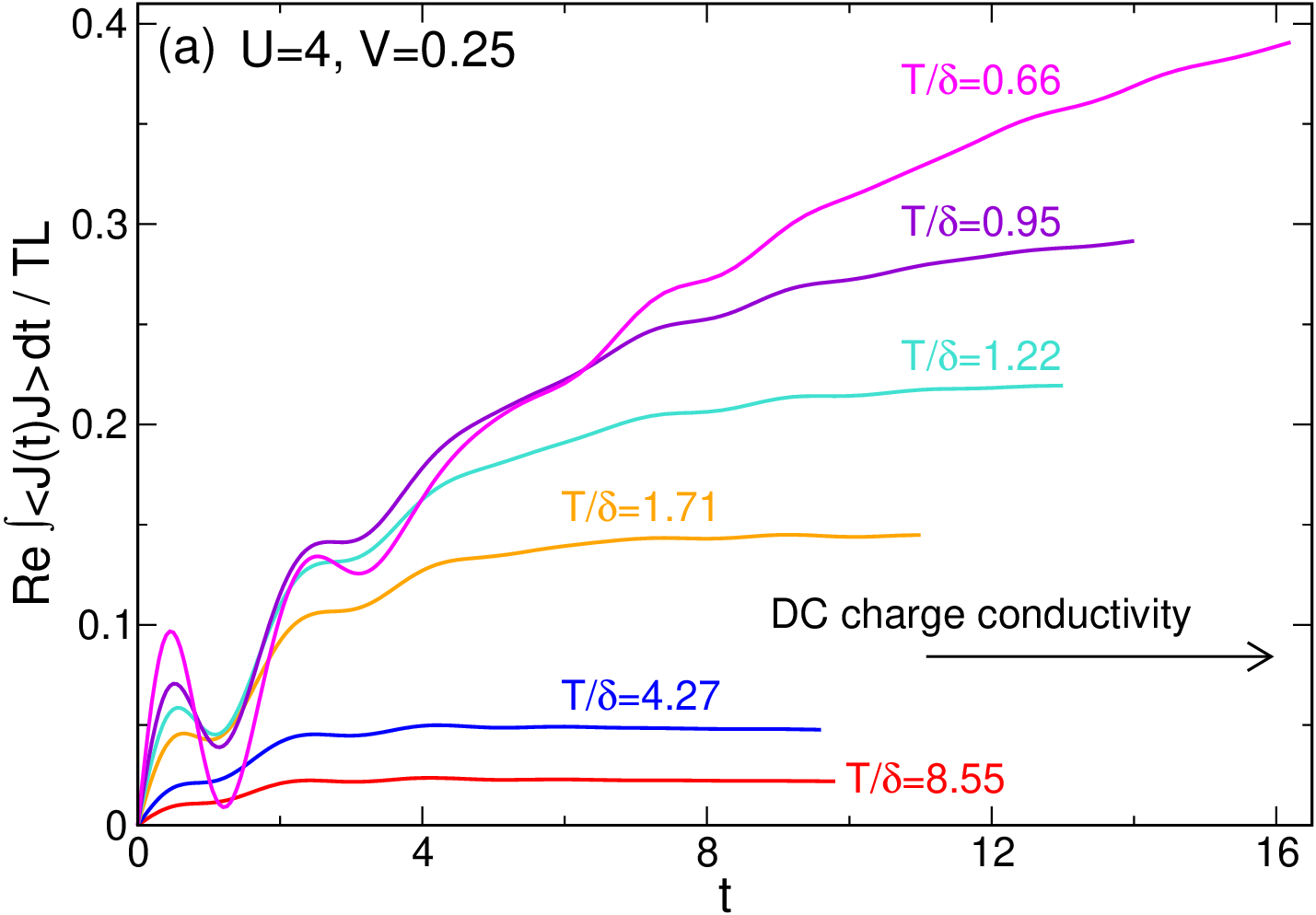}\hspace*{0.03\linewidth}
\includegraphics[width=0.475\linewidth,clip]{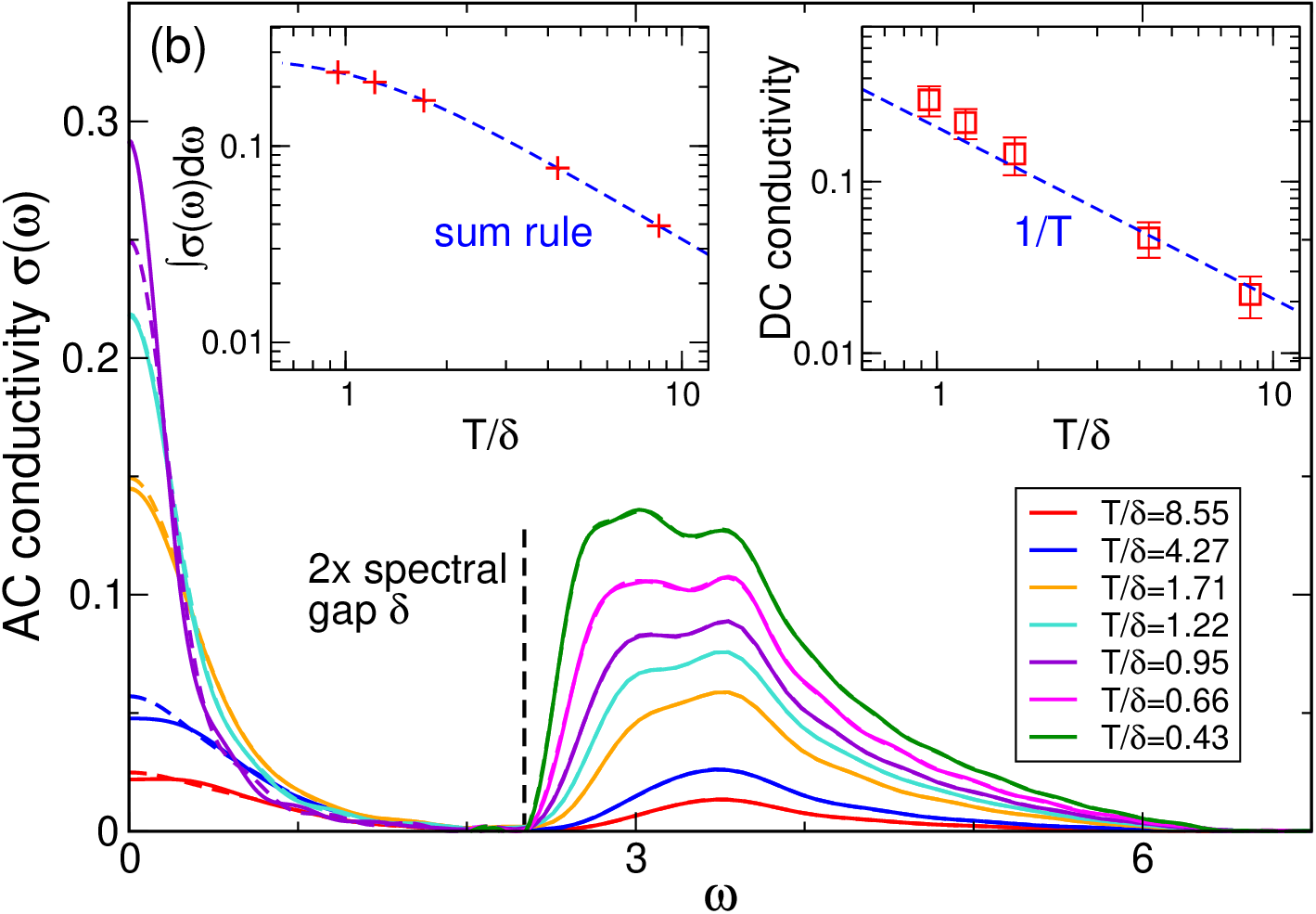}
\caption{(Color online) The same as in Fig.~\ref{fig:sigma_xxz} but for the Hubbard model where a potentially small finite Drude weight is eliminated by a small next-nearest neighbor interaction $V=U/16$ (which is far from the transition into another phase occurring around $V\approx U/2$\cite{extendedhub}). The DC conductivity is finite and increases if $T$ is lowered below the charge gap $\delta=1.17$ (measured at $V=0$; the optical gap is twice as large).\cite{hubbook} The curves at $T/\delta=0.66$ and $T/\delta=0.43$ are only shown for frequencies $\omega>2$. }
\label{fig:sigma_hub}
\end{figure*}

\section{Optical Conductivity}
\label{sec:cond}

\textit{XXZ chain} --- In this section we investigate the regular part of the optical conductivity. It is instructive to first study the simpler case of the XXZ chain in its gapped phase $\Delta>1$, particularly in order to illustrate how to assess the error when $\sigma_\tn{reg}(\omega)$ is computed from real time DMRG data. Fig.~\ref{fig:dw_xxz} again shows that the current correlation decays to zero for $\Delta>1$ and that thus the Drude weight vanishes. The DC conductivity is determined by the integral of $\langle J(t)J\rangle$ via Eq.~(\ref{eq:sigma1}). For anisotropies where the spectral gap $\delta$ is of order one, we can simulate up to times where this integral can be computed without having to resort to any extrapolation algorithms for all temperatures from $T\gg\delta$ to $T\ll\delta$ (we explain below how to estimate the error). This is illustrated in Fig.~\ref{fig:sigma_xxz}(a) where $\Delta=5$ and $\delta\sim3$. Interestingly, $\sigma_\tn{reg}(0)$ diverges even as $T$ is decreased below the gap [see the inset to Fig.~\ref{fig:sigma_xxz}(b)]. The temperature-dependence is approximately $\sigma_\tn{reg}(0)\sim1/T$, but we cannot rule out an additional prefactor that varies weakly with $T$. Such a divergence is consistent with a semiclassical analysis by Damle and Sachdev\cite{gaparg,gaparg2} who show that below the gap an exponentially small quasiparticle density is compensated by an exponentially long life time; for a model with spin-$1$ symmetry; this yields $\sigma_\tn{reg}(0)\sim1/\sqrt{T}$. However, to the best of our knowledge this picture was never confirmed conclusively for a microscopic model and all temperatures ranging from $T\gg\delta$ to $T\ll\delta$. Previous studies of the diffusion constant, which is related to $\sigma$ via an Einstein relation, can be found in Refs.~\onlinecite{diff_prosenznidaric,diff_steinigeweggemmer,diff_znidaric,diff_znidaric2,diff_steinigeweg2,finiteTquench} for $T=\infty$ as well as in Refs.~\onlinecite{diff_steinigeweg1,diff_znidaric2} for finite but large (compared to the gap) temperatures. Finally, we show the full frequency-dependent AC conductivity in Fig.~\ref{fig:sigma_xxz}(b). When the temperature is decreased, one observes two distinct features: a narrowing Lorentzian peak around $\omega=0$ as well as a sharp increase of $\sigma_\tn{reg}(\omega)$ at $\omega\sim\delta$.

Only finite times can be reached within the DMRG and it is thus essential to establish a controlled way to assess the associated error of $\sigma$. This can be achieved in various ways. First, one can compare the conductivities obtained via Eqs.~(\ref{eq:sigma1}) and (\ref{eq:sigma2}), which only coincide if the time integral extends to infinity, as exemplified by solid and dashed lines in Fig.~\ref{fig:sigma_xxz}(b). As expected, the finite-time error systematically becomes larger at small frequencies. Moreover, the missing contribution from larger times can be estimated by carrying out calculations with bigger discarded weights (which trades accuracy of the real-time data for reaching longer time scales) or by employing extrapolation schemes such as linear prediction (see Refs.~\onlinecite{barthel,scaling} for details). It seems reasonable to combine both strageties to define an error bar as twice as difference between $\sigma$ obtained with and without extrapolation or twice the difference between Eqs.~(\ref{eq:sigma1}) and (\ref{eq:sigma2}) (with extrapolation), whatever is larger. An example is shown in the right inset to Fig.~\ref{fig:sigma_xxz}(b). Finally, it is instructive is to verify the sum rule
\begin{equation}\label{eq:sumrule}
\int_0^\infty\sigma(\omega)d\omega=-\frac{\pi\langle H_\tn{kin}\rangle}{2L}\,, 
\end{equation}
where the kinetic energy $\langle H_\tn{kin}\rangle=\langle H(\Delta=0)\rangle$ can be obtained easily from infinite-system DMRG. The left inset to Fig.~\ref{fig:sigma_xxz}(b) illustrates that Eq.~(\ref{eq:sumrule}) is fulfilled with great accuracy.

\textit{Hubbard model} ---  We now turn to the Hubbard model. In order to rule out any subtleties due to a small but potentially finite Drude weight, we switch on a small next-nearest neighbor interaction $V=U/16$, which is far from the phase transition occurring at $V\approx U/2$.\cite{extendedhub} Fig.~\ref{fig:sigma_hub}(a) shows for $U=4$ (where the spectral gap is $\delta\sim1$)\cite{hubbook} that the charge current correlation functions decay to zero, and that their integral can be obtained for all temperatures from $T\gg\delta$ to $T\lesssim\delta$. At smaller $U\leq2$, the gap becomes exponentially small, and at low $T$ one can no longer reach time scales large enough to compute $\sigma_\tn{reg}(\omega)$ without using extrapolation schemes.

The optical charge conductivity of the almost-integrable Hubbard model is shown in Fig.~\ref{fig:sigma_hub}(b). As the temperature is decreased below the gap, a sharpening Lorentzian peak develops around $\omega\approx0$ -- the DC conductivity grows monotonously. Moreover, $\sigma_\tn{reg}(\omega)$ increases sharply for frequencies above the optical absorption threshold (which is twice as large as the charge gap\cite{hubbook}) and features a long tail that decays on a scale set by the width of the Hubbard bands ($4$ in our units). One successively approaches the zero temperature form of $\sigma_\tn{reg}(\omega)$ calculated in Ref.~\onlinecite{hubT0f} (see also Refs.~\onlinecite{hubT0b,hubT0d,hubT0e,hubT0g,hubT0h,hubT0i,hubT0j,hubT0k,hubT0l} for more results on $\sigma(\omega)$ at $T=0$; finite but small temperatures were considered in Ref.~\onlinecite{sigmahub}; an exact diagonalization study of small systems at $T=\infty$ can be found in Ref.~\onlinecite{edhub}; a lower bound on the infinite-temperature diffusion constant was established in Ref.~\onlinecite{prosenhub}).

\section{Summary and Outlook}

We studied finite-temperature linear response transport properties of the one-dimensional fermionic Hubbard model at half filling. Using real time DMRG numerics, we showed that the charge Drude weight is either zero or small; we established upper bounds. The optical charge conductivity was investigated in presence of small next-nearest neighbor interactions. Its DC part is finite and increases even as the temperature is lowered below the gap (the same holds for the integrable XXZ chain in its gapped phase). Our analysis thus suggests that the half-filled, gapped Hubbard model is a normal charge conductor at finite temperatures.

Aside from finding limits on the Drude weight and estimating transport properties of this long-standing model for comparison to experiment, we believe that the present work can serve as a starting point for considering the effects of additional perturbations, including those that break integrability. Our approach using time-dependent correlation functions can also be extended to compute nonequilibrium physics beyond linear response, as previously done for the XXZ model. Such nonequilibrium calculations could be compared to optical pump-probe experiments on correlated materials.  Finally, generalizing the calculations here to energy currents would allow calculations of the thermopower and the electronic contribution to thermal conductivity, which are particularly interesting in the context of  proposals to use conducting polymers as low-cost, flexible thermoelectric materials.

\emph{Acknowledgments} --- We are grateful to Fabian Heidrich-Meisner for useful comments. We acknowledge support by the Nanostructured Thermoelectrics program of LBNL (CK), by the Forschergruppe 723 of the DFG (DMK), and by AFOSR MURI as well as the Simons Foundation (JEM).


\end{document}